\documentclass[12pt,preprint]{aastex}

\newcommand{\rl}{$r$--$L$}

\shorttitle{The R--L Relationship: Host-Galaxy Starlight}
\shortauthors{Bentz, et al.}

\received{}
\accepted{}

\begin{document}

\title{The Radius--Luminosity Relationship for Active Galactic Nuclei:\\
 The Effect of Host-Galaxy Starlight on Luminosity Measurements}

\author{ Misty~C.~Bentz\altaffilmark{1}, Bradley~M.~Peterson\altaffilmark{1},
Richard~W.~Pogge\altaffilmark{1}, Marianne~Vestergaard\altaffilmark{2}, and 
Christopher~A.~Onken\altaffilmark{1,3}}

\altaffiltext{1}{Department of Astronomy, 
		The Ohio State University, 
		140 West 18th Avenue, 
		Columbus, OH 43210; 
		bentz, peterson, pogge@astronomy.ohio-state.edu}

\altaffiltext{2}{Steward Observatory, 
		The University of Arizona, 
		933 North Cherry Avenue, 
         	Tucson, AZ 85721; 
		mvestergaard@as.arizona.edu}

\altaffiltext{3}{Present address:
		National Research Council Canada, 
		Herzberg Institute of Astrophysics,
		5071 West Saanich Road,
		Victoria, BC  V9E 2E7;
		christopher.onken@nrc-cnrc.gc.ca }

\begin{abstract}

We have obtained high resolution images of the central regions of 14
reverberation-mapped active galactic nuclei (AGN) using the {\it Hubble
Space Telescope} Advanced Camera for Surveys High Resolution Camera to
account for host-galaxy starlight contamination of measured AGN
luminosities.  We measure the host-galaxy starlight contribution to the
continuum luminosity at 5100\,\AA\ through the typical ground-based slit
position and geometry used in the reverberation-mapping campaigns.  We
find that removing the starlight contribution results in a significant
correction to the luminosity of each AGN, both for lower luminosity
sources, as expected, but also for the higher luminosity sources such as
the PG quasars.  After accounting for the host galaxy starlight, we
revisit the well-known broad-line region radius--luminosity relationship
for nearby AGN.  We find the power-law slope of the relationship for the
H$\beta$ line to be $0.518 \pm 0.039$, shallower than previously
reported and consistent with the slope of $0.5$ expected from the naive
theoretical assumption that all AGN have, on average, the same ionizing
spectrum and the same ionization parameter and gas density in the
H$\beta$ line-emitting region.

\end{abstract}


\section{INTRODUCTION}

Over the past 15 years, reverberation-mapping
(\citealt{blandford82,peterson93}) efforts have led to measurements of
the size of the broad-line region (BLR), and consequently black hole
mass measurements, for 36 Seyfert~1 galaxies and low-luminosity quasars
(\citealt{peterson04,peterson05}).  Early efforts began with individual
objects and soon composed a handful of bright, relatively nearby Seyfert
galaxies (see \citealt{koratkar91a} and references therein for a summary
of early work).  Two very important relationships were anticipated and
searched for (e.g. \citealt{koratkar91b}): the radius--luminosity (\rl)
relationship (the relationship between the BLR size and optical
luminosity of the AGN) and the mass--luminosity relationship (the
relationship between the mass of the central black hole and the optical
luminosity).  While the \rl\ relationship was detectable in Seyfert 1
galaxies alone \citep{peterson93}, it was the addition of 17 quasars
from the Palomar-Green (PG) sample \citep{schmidt83} that allowed these
relationships to be studied in detail, as the larger range of
luminosities allowed for the emergence of a statistically significant
correlation \citep{kaspi00}.

With a calibrated \rl\ relationship, one can quickly estimate black hole
masses for large numbers of AGN with only two spectral measurements: the
velocity width of an emission line (usually H$\beta$) and the continuum
luminosity as a proxy for the radius.  This method of estimating masses
is especially useful for high redshift sources where
reverberation-mapping is impractical or impossible
(\citealt{wandel99,vestergaard02,vestergaard04,mclure02,wu04}; see also
\citealt{vestergaard06}).  Understanding the growth of black holes
throughout cosmic history is a key step to understanding the evolution
of galaxies and the universe, so correctly calibrating the \rl\
relationship is crucial.  In addition, a correct calibration is
necessary for understanding the physics of the central regions of AGN,
such as the ionization parameter, the ionizing spectral energy
distribution (SED), the gas density, and the column density.  These
parameters are inferred from the form of the \rl\ relationship, and our
understanding of them relies heavily on the accuracy with which the \rl\
relationship is known.

Recently, \citet{peterson04} compiled and consistently reanalyzed all
available re\-ver\-ber\-ation-mapping data for 35 AGN to improve their
black hole mass measurements.  Subsequently, the \rl\ relationship was
reexamined by \citet{kaspi05}.  Assuming a power-law relationship
$R_{BLR} \propto L^{\alpha}$, they find a best-fit value of $\alpha =
0.665 \pm 0.069$ using the optical continuum and broad H$\beta$ line.
However, some of the nearby AGN have host galaxy starlight luminosities
that are comparable to the luminosities of their central sources.  With
the relatively large apertures used during the AGN monitoring campaigns
(typically on the order of 5$\farcs$0 $\times$ 7$\farcs$5), the
host-galaxy starlight contribution is substantial.  Ignoring the
starlight contribution to the optical luminosity will overestimate the
optical luminosity of the source and artificially steepen the slope of
the \rl\ relationship, as the starlight fraction is relatively more
important in the lower luminosity objects.

Previous attempts to quantify the host-galaxy starlight contribution to
the optical flux at 5100\,\AA\ were carried out using ground-based
telescopes with $1 - 2 \arcsec$ seeing
(\citealt{alloin95,stirpe94,peterson95,romanishin95}).  However, the
relatively low resolution available from the ground of even the most
nearby sources, coupled with seeing effects on the order of $1 \arcsec$,
make it almost impossible to disentangle the nucleus and the bulge of
the galaxy.  In this paper, we use high resolution {\it Hubble Space
Telescope} (HST) images to measure the contribution from starlight to
previous luminosity measurements of 14 reverberation-mapped AGN and we
present a revised \rl\ relationship in which these effects are taken
into account.

\section{OBSERVATIONS AND DATA REDUCTION}
Between 2003 August 22 and 2004 May 28, we observed 14 Seyfert~1
galaxies with the {\it HST} Advanced Camera for Surveys (ACS) in
Snapshot mode.  The targets and details of the observations are listed
in Table~1.  Each object was imaged with the High Resolution Camera
(HRC) through the F550M filter ($\lambda_{c} = 5580$\,\AA\ and $\Delta
\lambda = 547$\,\AA), thereby probing the continuum while avoiding
strong emission lines.  The observations consisted of three exposures
for each object with exposure times of 120\,s, 300\,s, and 600\,s.  This
method of graduating the exposure times was employed to avoid saturation
of the nucleus but still obtain a reasonable signal-to-noise ratio
($S/N$) for the wings of the point-spread function (PSF) and the host
galaxy.  Each individual exposure was split into two equal sub-exposures
to facilitate the rejection of cosmic rays (see
e.g. \citealt{carollo97,ho2001} for discussions of the challenges
involved in obtaining accurate optical photometric data of faint nuclei
in bright nearby galaxies).

The nuclei of most of the 600\,s exposures were saturated, and the
nuclei of the 300\,s observations of NGC\,3783 and PG\,0844+349 were
also saturated.  Only 3C\,390.3, Fairall\,9, Mrk\,110, Mrk\,590, and
NGC\,5548 were unsaturated in all three graduated exposures.

The data quality frames provided by the {\it HST} pipeline were
consulted to identify the individual saturated pixels associated with
the nucleus in each exposure frame.  These saturated pixels were clipped
from the image and replaced by the same pixels from a non-saturated
exposure after scaling them by the relative exposure times.  The three
frames for each object were then summed to give one frame with an
effective exposure time of 1020\,s for each of the 14 objects.

Cosmic rays were identified in the summed images with the Laplacian
cosmic ray identification package L.A.Cosmic \citep{vandokkum01}.
Pixels in the PSF area of each image that were identified by L.A.Cosmic
were excluded from the list of affected pixels prior to cleaning with
XVista.\footnote{XVISTA was originally developed as Lick Observatory
Vista and is now maintained in the public domain by former Lick graduate
students as a service to the community.  It is currently maintained by
Jon Holtzman at New Mexico State University, and is available at
http://ganymede.nmsu.edu/holtz/xvista.}  Each remaining affected pixel
was replaced with the median value for the eight pixels immediately
surrounding it.

Finally, the summed, cleaned images were corrected for the distortions
of the ACS camera with the PyRAF routine {\it pydrizzle} in the
STSDAS\footnote{STSDAS and PyRAF are products of the Space Telescope
Science Institute, which is operated by AURA for NASA.} package for
IRAF\footnote{IRAF is distributed by the National Optical Astronomical
Observatory, which is operated by the Association of Universities for
Research in Astronomy, Inc., under cooperative agreement with the NSF.}.

The final images for four representative galaxies --- 3C\,120,
3C\,390.3, NGC\,3227, and NGC\,5548 --- are shown in Figure~1, overlaid
with the apertures that were used in their ground-based monitoring
campaigns.

\section{GALAXY DECOMPOSITION}
Each object was fit with typical spiral galaxy parameters in order to
determine and accurately subtract the contribution from the central
point source.  The galaxies were modeled using the two-dimensional image
decomposition program Galfit \citep{peng02}, which fits analytic
functions for the bulge and disk, plus an additional point source for
the nucleus, convolved with a user-supplied model PSF.  The simulated
PSF was created using the TinyTim package \citep{krist93}, which models
the optics of {\it HST} plus the specifics of the camera and filter
system.

Several initial attempts to fit the galaxies led to the following
relatively robust models for the individual components of each galaxy:
bulges were fit with a \citet{devaucouleurs48} $R^{1/4}$ profile, disks
were fit with an exponential model, and the central PSF was created
using TinyTim as discussed above.  Each parameter governing the models
was allowed to vary from the initial conditions except the boxy/disky
parameter, which sometimes led to instabilities in the fits and was
therefore held fixed at zero for each of the fits.  Figure 2 shows the
galaxy fits and residuals for the four representative objects ---
3C\,120, 3C\,390.3, NGC\,3227, and NGC\,5548 --- spanning the range of
fit quality encountered for the various angular sizes of objects in our
sample.

Once the fits were relatively stable from perturbations on the initial
conditions, the best-fit central PSFs were subtracted from the images,
resulting in a nucleus-free image of each of the 14 objects, which are
shown in Figure~3.  Below, we describe the detailed results of the fits
to individual galaxies.

\paragraph{3C\,120.}

A central PSF, a deVaucouleurs profile, and an exponential disk were
fitted to 3C\,120, resulting in a rather clean subtraction.  A tidal
tail west of the nucleus and trailing to the north remains in the
residuals (see Figure~2).

\paragraph{3C\,390.3.}

We fit 3C\,390.3 with a central PSF, a deVaucouleurs profile, and an
exponential disk.  The residuals are the cleanest for any of the 14
objects in this particular sample and reveal no evidence for any
underlying structure (see Figure~2).

\paragraph{Fairall\,9.}

A central PSF, a deVaucouleurs profile, and an exponential disk was fit
to Fairall\,9.  A strong bar oriented in the east--west direction is
clearly evident in the fit residuals for Fairall~9.

\paragraph{Markarian\,110.}

The fit for Markarian\,110 included only a central PSF and a
deVaucouleurs profile, as the disk surface brightness was evidently too
low to be detected in these observations.  The apparent double nucleus,
the result of a foreground star superimposed on the galaxy to the
northeast of the central source \citep{hutchings88}, is clearly resolved
in the images, and the residuals are very clean with no evidence for any
additional underlying structure.

\paragraph{Markarian\,279.}

We fit Markarian\,279 with a central PSF, a deVaucouleurs profile, and
an exponential disk.  Faint spiral arms are visible in the fit
residuals.

\paragraph{Markarian\,590.}

Only a central PSF and a deVaucouleurs profile were fit to
Markarian\,590.  There appears to be a rather bright bar in the center
of the galaxy on a scale of approximately 3--4\arcsec\ and the spiral
arms appear to penetrate all the way to the very nucleus of the galaxy,
although whether the inner spiral arms are connected to the outer spiral
arms is unclear.

\paragraph{Markarian\,817.}

We fit Markarian\,817 with a central PSF, a deVaucouleurs profile, and
an exponential disk.  The residuals clearly show the central bar and two
tightly wound spiral arms.

\paragraph{NGC\,3227.}

The fit for NGC\,3227 was one of the most marginal fits we encountered.
It included a central PSF, a deVaucouleurs profile, and an exponential
disk.  The large amounts of dust in the center of the galaxy make for an
interesting residual image (see Figure~2).

\paragraph{NGC\,3783.}

A central PSF, a deVaucouleurs profile, and an exponential disk were fit
to NGC\,3783.  There appears to be faint evidence for a bar at a position angle (PA) of roughly $160 \degr$.

\paragraph{NGC\,4051.}

We fit NGC\,4051 with a central PSF, a deVaucouleurs profile, and an
exponential disk.  The residuals show a bright circumnuclear ring as
well as areas of dust absorption.

\paragraph{NGC\,4151.}

The fit for NGC\,4151 included a central PSF, a deVaucouleurs profile,
and an exponential disk.  The residuals show a fairly messy central
region with dust absorption and areas of excess luminosity.

\paragraph{NGC\,5548.}

A central PSF, a deVaucouleurs profile, and an exponential disk were fit
to NGC\,5548.  The spiral arms are clearly visible in the residuals, as
well as various knots of star formation in the arms (see Figure~2).

\paragraph{PG\,0844+349.}

We fit PG\,0844+349 with a central PSF, a deVaucouleurs profile, and an
exponential disk.  The residuals are very clean and show no evidence for
underlying structure.

\paragraph{PG\,2130+099.}

Only a central PSF and a deVaucouleurs profile were necessary to fit
PG\,2130+099.  The residuals are fairly clean but hint at a tidal tail
trailing towards the west.

\section{FLUX MEASUREMENTS}

The nucleus-free image of each galaxy was overlaid with the typical
aperture used in its ground-based monitoring program at the typical
orientation and centered on the position of the AGN (see Table~2).  The
counts within the aperture were summed and converted to $f_{\lambda}$
flux density units (erg s$^{-1}$ cm$^{-2}$ \AA$^{-1}$) using the {\it
HST} keyword PHOTFLAM and the effective exposure time for each object.

Color corrections between the flux observed through the {\it HST} F550M
filter and restframe 5100\,\AA\ were calculated using a model bulge
spectrum \citep{kinney96} plus a powerlaw component of the form $f_{\nu}
\propto \nu^{-0.5}$ for the AGN component.  The relative amounts of
emission from each of the two components were set by the measured
fractions of bulge and AGN emission in each combined {\it HST} image.
All of the color corrections were close to one.  Following
\citet{kaspi05}, we corrected for galactic absorption using the
\citet{schlegel98} $E(B-V)$ values listed in the NASA/IPAC Extragalactic
Database (NED) and the extinction curve of \citet{cardelli89}, adjusted
to $A_V/E(B-V) = 3.1$. Table~3 lists the color correction for each
galaxy as the ratio of the flux at restframe 5100~\AA\ to the average
flux through the {\it HST} F550M filter, as well as the final
host-galaxy flux measurement at 5100 \AA\ for each galaxy through the
monitoring aperture listed in Table~2.

It is interesting to note that these galaxy flux measurements are not in
good agreement with previous measurements made from the ground in a
similar fashion (see Table~3).  Our own ground-based tests with the
1.3-m McGraw-Hill Telescope at MDM Observatory revealed that the typical
resolution achieved from the ground coupled with the 1-2~\arcsec\ seeing
typical of ground-based projects resulted in the PSF of the central
source and the bulge smearing together in an indistinguishable fashion.
Our attempts to fit the galaxy with ground-based images were more
dependent on the initial parameters supplied to Galfit than on any
information in the images themselves.  Therefore, we strongly recommend
that future projects rely only on high spatial resolution images such as
those acquired with space-based or diffraction-limited telescopes.

\section{THE RADIUS-LUMINOSITY RELATIONSHIP}

We have calculated several types of fits to the \rl\ relationship for
the 35 reverberation-mapped AGN after correcting the above 14 for
starlight from the host galaxy.  These calculations will be refined in
the future as we correct additional reverberation-mapped AGN for
host-galaxy starlight contributions.  Following \citet{kaspi05}, the
\rl\ fits have been calculated for the H$\beta$ line only as well as for
the Balmer-line average.  Within these divisions, we have also made the
distinction of treating each separate measurement of an object
individually, as well as taking the mean of multiple measurements
weighted by the average of the positive and negative errors.  We tested
the differences between weighting measurements by the average of their
errors, by taking only the positive errors, and by taking the errors
toward the fit in the manner of \citet{kaspi05}.  We find the
differences in these weighting methods to be at the 2$\%$ level, and
therefore negligible.

For those fits that used only the H$\beta$ BLR radius, three objects
(PG\,0844+349, PG\,1211+143, and NGC\,4593) were determined by
\citet{peterson04} to have unreliable H$\beta$ time delays and these
objects are therefore excluded from any fits to the H$\beta$ \rl\
relationship.  Other fits excluded NGC\,3516, IC\,4329A, and NGC\,7469
because they have a significant, but unquantified, host-galaxy starlight
contribution that we are unable to correct for at this time.\footnote{We
have an approved {\it HST} Cycle 14 program to quantify the starlight
contribution from these objects in the same manner as the objects
examined in this work.}  For some fits, we also excluded NGC\,3227 and
NGC\,4051, as they are well known to have significant nuclear structure
and reddening, as well as PG\,2130+099, which is a clear outlier and for
which we now believe the radius measurement is probably erroneous (this
will be discussed elsewhere in more detail).

We have used three different methods to calculate the relationship
between the size of the BLR and the optical luminosity:

\begin{description}
\item[1.] FITEXY \citep{press92}, which estimates the parameters of a 
        straight-line fit through the data including errors in both
        coordinates.  FITEXY numerically solves for the minimum
        orthogonal $\chi^2$ using an interative root-finding algorithm.
        We include intrinsic scatter similar to \citet{kaspi05}.
        Namely, the fractional scatter listed in Table~5 is the fraction
        of the measurement value of $r$ (not the error value) that is
        added in quadrature to the error value so as to obtain a reduced
        $\chi^2$ of 1.0.
\item[2.] BCES  \citep{akritas96}, which attempts to account for the 
        effects of errors on both coordinates in the fit using bivariate
        correlated errors, including a component of intrinsic scatter.
        We adopt the bootstrap of the bisector value following
        \citet{kaspi05}.
\item[3.] GaussFit \citep{mcarthur94}, which implements generalized 
	least-squares using robust Householder Orthogonal
	Transformations \citep{jefferys80,jefferys81} to solve the
	non-linear equations of condition for the problem of errors in
	both coordinates.
\end{description}

Table~4 lists the previous luminosity measurement \citep{kaspi05}, the
luminosity after correction for host-galaxy starlight, and the H$\beta$
time lag data \citep{peterson04} for each of the 14 reverberation-mapped
AGN in this study.  For the remaining 21 reverberation-mapped AGN that
were not corrected in this work, their H$\beta$ time lags can be found
in Table~6 of \citet{peterson04} and their luminosities and Balmer-line
averaged time lags are available in Table~1 of \citet{kaspi05}.

Table~5 lists the various fits to the \rl\ relationship for each of the
fitting methods discussed above.  The fit parameters listed in Table~5
are appropriate for fits to the function:

\begin{equation}
{\rm log}\,(R_{\rm BLR}) = K + \alpha\ {\rm log}\,(\lambda L_{\lambda}\,(5100 \rm\AA))
\end{equation}
 
\noindent
where $\alpha$ is the slope of the power-law relationship between $R_{\rm
BLR}$ and $L$\,(5100\,\AA) and $K$ is the scaling factor.

It is easily observed that the calculated slope of the \rl\ relationship
in Table~5 does not depend sensitively on the fitting method employed or
the inclusion or exclusion of certain suspect data points.  For the
remainder of this paper we will adopt the best-fit parameters calculated
using GaussFit, shown in bold face in Table~5 and plotted in Figure~4
against the best fit determined by \citet{kaspi05}.  The generalized
least-squares method of \citet{jefferys80,jefferys81} implemented by
GaussFit has the virtue of directly addressing the explicit
non-linearity of the problem of fitting lines through data with errors
in both variables, whereas BCES and FITEXY use iterative linear
approximations to estimate the best-fit parameters.  We used these
latter two methods only because they were employed by \citet{kaspi05},
and this allows us to make a more direct comparison with their results.

\section{DISCUSSION}

The first models of the BLRs of AGN were modified from planetary nebulae
models, as the emitting gas in the first studied AGN looked somewhat
similar to the clouds or filaments of nebulae \citep{greenstein64}.
This typical BLR model of optically thick, line-emitting gas did not
change significantly for several years (see the review by
\citealt{davidson79}).  Eventually, various other models of the BLR were
added.  They included cool clouds embedded in hot gas \citep{krolik81},
bloated stars (\citealt{scoville88,kazanas89,alexander94}),
magnetically-driven disk winds \citep{emmering92}, and the combination
of disk and disk wind components \citep{murray95}.

Photoionization equilibrium codes have often been employed to study the
basic characteristics of the BLR as observed through the AGN spectra
(early versions: \citealt{davidson77,davidson79,kwan81}; later versions:
\citealt{rees89,goad93,kaspi99,baldwin95,korista00}).  Typically, the
line-emitting gas clouds are parameterized by elemental abundance, the
shape of the ionizing continuum, and an ionization parameter

\begin{equation}
U = \frac{Q(H)}{4 \pi r^2 c n_{e}}\ ,
\end{equation}

\noindent 
where

\begin{equation}
Q(H) = \int^{\infty}_{\nu_{1}} \frac{L_{\nu}}{h \nu}\, d\nu
\end{equation}

\noindent 
is the number of photons with energies in excess of $h \nu_{1} = 13.6\
{\rm eV}$ (the energy required to ionize hydrogen) emitted each second
by the central source.

To the lowest order, all AGN spectra are remarkably
similar.\footnote{For a visual representation of this remarkable fact,
see the luminosity-binned composite spectra from the Sloan Digital Sky
Survey presented by \citet{vandenberk04}.  The largest differences
between the composites are due to the ``Baldwin Effect'' --- the
\ion{C}{4} $\lambda 1549$ emission line decreases in equivalent width as
the continuum luminosity increases \citep{baldwin77}.}  Therefore, one
can assume that the ionization parameters and particle densities are
about the same for all AGN.  Rearranging the definition of the
ionization parameter $U$, we find:

\begin{equation}
r = \left( \frac{Q(H)}{4 \pi c n_{e} U} \right) ^{1/2} \propto Q(H)^{1/2}.
\end{equation}

\noindent
If we further assume that the shape of the ionizing continuum is not a
function of luminosity, then $L \propto Q(H)$ so that

\begin{equation}
r \propto L^{1/2}
\end{equation}

\noindent
and therefore we can expect the slope of the \rl\ relationship to be
$\alpha \approx 0.5$.

Accounting for host-galaxy starlight emission in the 14 objects with
high resolution {\it HST} imaging, we find a significant deviation from
the \rl\ fit parameters published by \citet{kaspi05}.  Figure~4 shows
both the previous \rl\ fit found by \citet{kaspi05} as well as the new
fit that we present here.  Comparing the H$\beta$ \rl\ fit, where
multiple measurements are averaged together, \citet{kaspi05} quote a
power-law slope of $\alpha = 0.665 \pm 0.069$, while we find a power-law
slope of $\alpha = 0.518 \pm 0.039$ when excluding any sources with
significant but unquantified host galaxy emission or strong reddening.
Our value of the slope is in good agreement with the expectation of
$\alpha = 0.5$, based on the naive assumptions that the ionization
parameters, gas densities, and ionizing SEDs of all AGNs are the same.

\section{CONCLUSIONS}

We have measured the host-galaxy starlight component to previous
luminosity measurements of 14 reverberation-mapped AGN through their
original monitoring apertures.  Removing the starlight component from
the luminosity measurements, we recalculate the \rl\ relationship for
nearby AGN.  Regardless of the detailed selection of the sample or the
regression method used to determine the fit, we find a slope of $\alpha
\approx 0.50$ for H$\beta$ and $\alpha \approx 0.51$ for the mean of the
Balmer lines, both consistent with the naive photoionization expectation
of $\alpha = 0.5$ if all AGN have the same ionization parameter, gas
density, and ionizing SED.

Additional high-resolution optical imaging by {\it HST} is necessary to
constrain the amount of host-galaxy starlight contribution to the
remaining population of reverberation-mapped AGN, and we are therefore
continuing this investigation.

Better constraints on the H$\beta$ BLR radius are also necessary for
many of these objects and additional reverberation experiments are being
undertaken.

In order to constrain the true nature of the relationship between the
BLR radius and luminosity of the bright AGN in the local universe, each
of the objects contributing to the fit of this relationship must be
studied and understood in as much detail as possible.  A thorough
understanding of the local form of this relationship, especially its
biases or weaknesses, is crucial if we are to apply it in the hopes of
understanding the growth and evolution of AGN and galaxies from the
early universe to the present day.

\acknowledgements
We would like to thank an anonymous referee for comments that improved
the presentation of this paper.  This work is based on observations with
the NASA/ESA {\it Hubble Space Telescope}.  We are grateful for support
of this work through grant {\it HST} GO-9851 from the Space Telescope
Science Institute, which is operated by the Association of Universities
for Research in Astronomy, Inc., under NASA contract NAS5-26555, and by
the NSF through grant AST-0205964 to The Ohio State University.  M.B. is
supported by a Graduate Fellowship from the National Science Foundation.
M.V. acknowledges financial support from NSF grant AST-0307384 to the
University of Arizona.  This research has made use of the NASA/IPAC
Extragalactic Database (NED) which is operated by the Jet Propulsion
Laboratory, California Institute of Technology, under contract with the
National Aeronautics and Space Administration and the SIMBAD database,
operated at CDS, Strasbourg, France.

\clearpage

\begin{deluxetable}{lcccccc}
\tablecolumns{7}
\tablewidth{0pt}
\tabletypesize{\small}
\tablecaption{Observation Log}
\tablehead{
\colhead{Objects} &
\colhead{References\tablenotemark{a}} &
\colhead{$z$} &
\colhead{$D_{L}$} &
\colhead{Date Observed} &
\colhead{Beginning UTC} &
\colhead{Datasets}\\
\colhead{} &
\colhead{} &
\colhead{} &
\colhead{(Mpc)} &
\colhead{(yyyy--mm--dd)} &
\colhead{(hh:mm:ss)} &
\colhead{(J8SC)}}

\startdata
Fairall\,9   & 1     & 0.04702 & 209 & 2003-08-22 & 00:44:00 & 04011,04021,04031 \\
Mrk\,590     & 2     & 0.02639 & 115 & 2003-12-18 & 02:27:00 & 05011,05021,05031 \\
3C\,120	     & 2     & 0.03301 & 145 & 2003-12-05 & 05:48:00 & 06011,06021,06031 \\
PG\,0844+349 & 3     & 0.06400 & 287 & 2004-05-10 & 20:11:00 & 10011,10021,10031 \\
Mrk\,110     & 2     & 0.03529 & 155 & 2004-05-28 & 17:34:00 & 11011,11021,11031 \\
NGC\,3227    & 4,5,6 & 0.00386 & 17  & 2004-03-20 & 04:28:00 & 13011,13021,13031 \\
NGC\,3783    & 7,8   & 0.00973 & 42  & 2003-11-15 & 00:11:00 & 15011,15021,15031 \\
NGC\,4051    & 9     & 0.00234 & 10  & 2004-02-16 & 01:49:00 & 16011,16021,16031 \\
NGC\,4151    & 10,11 & 0.00332 & 14  & 2004-03-28 & 14:25:00 & 17011,17021,17031 \\
Mrk\,279     & 12,13 & 0.03045 & 133 & 2003-12-07 & 03:54:00 & 24011,24021,24031 \\
NGC\,5548    & 14,15 & 0.01718 & 75  & 2004-04-07 & 01:53:00 & 27011,27021,27031 \\
Mrk\,817     & 2     & 0.03146 & 138 & 2003-12-08 & 18:08:00 & 29011,29021,29031 \\
3C\,390.3    & 16    & 0.05610 & 251 & 2004-03-31 & 06:56:00 & 34011,34021,34031 \\
PG\,2130+099 & 3     & 0.06298 & 283 & 2003-10-21 & 06:47:00 & 36011,36021,36031 \\
\enddata
\tablenotetext{a}{References refer to reverberation-mapping campaigns in 
                  optical wavelengths.}
\tablerefs{1. \citet{santoslleo97},
	   2. \citet{peterson98},
	   3. \citet{kaspi00},
	   4. \citet{salamanca94},
	   5. \citet{onken03},
	   6. \citet{winge95},
	   7. \citet{stirpe94},
	   8. \citet{onken02},
	   9. \citet{peterson00},
          10. \citet{kaspi96},
	  11. \citet{maoz91},
	  12. \citet{santoslleo01},
	  13. \citet{maoz90},
	  14. \citet{peterson02} and references therein,
	  15. \citet{netzer90},
	  16. \citet{dietrich98},
}
\end{deluxetable}

\clearpage

\begin{deluxetable}{lcccc}
\tablecolumns{4}
\tablewidth{0pt}
\tablecaption{Ground-Based Monitoring Aperture Sizes and Orientations}
\tablehead{
\colhead{Object} &
\colhead{PA ($\degr$)} &
\colhead{Aperture ($\arcsec$)} &
\colhead{References}}
\startdata
Fairall\,9   & 0     & $4 \times 9$   & 1 \\
Mrk\,590     & 90    & $5 \times 7.6$ & 2 \\
3C\,120      & 90    & $5 \times 7.6$ & 2 \\
PG\,0844+349 & 36.8  & $10 \times 13$ & 3 \\
Mrk\,110     & 90    & $5 \times 7.6$ & 2 \\
NGC\,3227    & 25    & $1.5 \times 4$ & 4 \\
NGC\,3783    & 0     & $5 \times 10$  & 5 \\
NGC\,4051    & 90    & $5 \times 7.5$ & 6 \\
NGC\,4151    & 156.3 & $10 \times 13$ & 7 \\
Mrk\,279     & 90    & $5 \times 7.5$ & 8 \\ 
NGC\,5548    & 90    & $5 \times 7.5$ & 9 \\
Mrk\,817     & 90    & $5 \times 7.6$ & 2 \\
3C\,390.3    & 90    & $5 \times 7.5$ & 10 \\
PG\,2130+099 & 68.2  & $10 \times 13$ & 3 \\
\enddata
\tablerefs{1. \citet{santoslleo97},
	   2. \citet{peterson98}, 
           3. \citet{kaspi00},
           4. \citet{salamanca94},
	   5. \citet{stirpe94}, 
           6. \citet{peterson00},
	   7. \citet{kaspi96},
	   8. \citet{santoslleo01},
	   9. \citet{peterson02} and references therein,
	  10. \citet{dietrich98}.
}
\end{deluxetable}

\clearpage

\begin{deluxetable}{lcccc}
\tablecolumns{5}
\tablewidth{0pt}
\tabletypesize{\small}
\tablecaption{Galaxy Flux}
\tablehead{
\colhead{Object} &
\colhead{Color Term} &
\colhead{$f_{gal} (5100 {\rm \AA})$ \tablenotemark{a}} &
\colhead{$f_{GB} (5100 {\rm \AA})$ \tablenotemark{b}} &
\colhead{References} \\
\colhead{} & 
\colhead{$f_{5100 {\rm \AA}} /  f_{F550M}$} & 
\colhead{($10^{-15}$ erg s$^{-1}$ cm$^{-2}$ \AA$^{-1}$)} &
\colhead{($10^{-15}$ erg s$^{-1}$ cm$^{-2}$ \AA$^{-1}$)} &
\colhead{}}

\startdata
Fairall\,9   & 1.002 & 4.32$^{+0.73}_{-0.87}$   & & \\
Mrk\,590     & 0.953 & 4.83$^{+0.81}_{-0.98}$   & & \\
3C\,120      & 1.046 & 1.82$^{+0.31}_{-0.37}$   & & \\
PG\,0844+349 & 1.018 & 2.66$^{+0.45}_{-0.54}$   & & \\
Mrk\,110     & 1.076 & 1.11$^{+0.19}_{-0.23}$   & & \\
NGC\,3227    & 1.009 & 5.96$^{+1.00}_{-1.21}$   & & \\
NGC\,3783    & 1.027 & 7.60$^{+1.28}_{-1.54}$   & 10.99 & 1,2 \\
NGC\,4051    & 0.987 & 10.43$^{+1.75}_{-2.11}$  & & \\
NGC\,4151    & 0.960 & 29.77$^{+5.01}_{-6.02}$  & 14.0 & 3 \\
Mrk\,279     & 1.031 & 3.68$^{+0.62}_{-0.75}$   & & \\
NGC\,5548    & 0.965 & 4.47$^{+0.75}_{-0.90}$   & 3.4 & 3,4 \\
Mrk\,817     & 1.030 & 2.48$^{+0.42}_{-0.50}$   & & \\
3C\,390.3    & 1.018 & 1.19$^{+0.20}_{-0.24}$   & & \\
PG\,2130+099 & 1.021 & 1.91$^{+0.32}_{-0.39}$   & & \\
\enddata
\tablenotetext{a} {Galaxy fluxes through the apertures described in 
		   Table~2 after subtraction of the central sources and
		   color corrections from the {\it HST} F550M filter to
		   5100\,\AA.  Fluxes are not corrected for galactic
		   extinction.}
\tablenotetext{b} {Galaxy fluxes in the literature determined from 
                   ground-based images.}
\tablerefs{1. \citet{alloin95}, 
	   2. \citet{stirpe94}, 
           3. \citet{peterson95},
	   4. \citet{romanishin95}. }

\end{deluxetable}

\clearpage

\begin{deluxetable}{lccc}
\tablecolumns{4}
\tablewidth{0pt}
\tablecaption{Optical Luminosities and H$\beta$ BLR Radii}
\tablehead{
\colhead{Object} &
\colhead{Previous $\lambda L_{\lambda}$(5100\,\AA)\tablenotemark{a}} &
\colhead{Corrected $\lambda L_{\lambda}$(5100\,\AA)\tablenotemark{b}} &
\colhead{R$_{BLR}$\tablenotemark{c}} \\
\colhead{} &
\colhead{(10$^{44}$ erg s$^{-1}$)} &
\colhead{(10$^{44}$ erg s$^{-1}$)} &
\colhead{(lt days)}}
\startdata
Fairall\,9   & 1.79 $\pm$ 0.20 		& 0.49 $\pm$ 0.25 	& 17.4$^{+3.2}_{-4.3}$ \\

Mrk\,590     & 0.736 $\pm$ 0.058 	& 0.288 $\pm$ 0.066	& 20.7$^{+3.5}_{-2.7}$ \\
	     & 0.497 $\pm$ 0.053 	& 0.0468 $\pm$ 0.0978 	& 14.0$^{+8.5}_{-8.8}$ \\
	     & 0.594 $\pm$ 0.042	& 0.145 $\pm$ 0.050 	& 29.2$^{+4.9}_{-5.0}$ \\
	     & 0.786 $\pm$ 0.122 	& 0.339 $\pm$ 0.151 	& 28.8$^{+3.6}_{-4.2}$ \\

3C\,120      & 1.39 $\pm$ 0.25 		& 0.851 $\pm$ 0.324 	& 38.1$^{+21.3}_{-15.3}$ \\

PG\,0844+349 & 2.21 $\pm$ 0.23 		& 0.631 $\pm$ 0.281 	& 3.0$^{+12.4}_{-10.0}$ \tablenotemark{d} \\

Mrk\,110     & 0.547 $\pm$ 0.057 	& 0.372 $\pm$ 0.065 	& 24.3$^{+5.5}_{-8.3}$ \\
             & 0.628 $\pm$ 0.080 	& 0.454 $\pm$ 0.086 	& 20.4$^{+10.5}_{-6.3}$ \\
             & 0.42 $\pm$ 0.14 		& 0.246 $\pm$ 0.182 	& 33.3$^{+14.9}_{-10.0}$ \\

NGC\,3227    & 0.0256 $\pm$ 0.0044 	& 0.0151 $\pm$ 0.0053 	& 8.2$^{+5.1}_{-8.4}$ \\
	
NGC\,3783    & 0.178 $\pm$ 0.015 	& 0.0603 $\pm$ 0.0174 	& 10.2$^{+3.3}_{-2.3}$ \\

NGC\,4051    & 0.0086 $\pm$ 0.0006 	& 0.00191 $\pm$ 0.00067 & 5.8$^{+2.6}_{-1.8}$ \\

NGC\,4151    & 0.1110 $\pm$ 0.0064 	& 0.0708 $\pm$ 0.0068 	& 3.1$^{+1.3}_{-1.3}$ \\

Mrk\,279     & 0.810 $\pm$ 0.082 	& 0.383 $\pm$ 0.088 	& 16.7$^{+3.9}_{-3.9}$ \\
		
NGC\,5548    & 0.362 $\pm$ 0.046 	& 0.200 $\pm$ 0.052 	& 19.7$^{+1.5}_{-1.5}$ \\
	     & 0.260 $\pm$ 0.037 	& 0.102 $\pm$ 0.046 	& 18.6$^{+2.1}_{-2.3}$ \\
	     & 0.343 $\pm$ 0.034 	& 0.182 $\pm$ 0.037 	& 15.9$^{+2.9}_{-2.5}$ \\
	     & 0.246 $\pm$ 0.043 	& 0.0832 $\pm$ 0.0549 	& 11.0$^{+1.9}_{-2.0}$ \\
	     & 0.331 $\pm$ 0.032 	& 0.170 $\pm$ 0.034 	& 13.0$^{+1.6}_{-1.4}$ \\
	     & 0.356 $\pm$ 0.040 	& 0.195 $\pm$ 0.045 	& 13.4$^{+3.8}_{-4.3}$ \\
	     & 0.442 $\pm$ 0.037 	& 0.282 $\pm$ 0.042 	& 21.7$^{+2.6}_{-2.6}$ \\
	     & 0.386 $\pm$ 0.060 	& 0.2224 $\pm$ 0.071 	& 16.4$^{+1.2}_{-1.1}$ \\
	     & 0.297 $\pm$ 0.033 	& 0.135 $\pm$ 0.039 	& 17.5$^{+2.0}_{-1.6}$ \\
	     & 0.492 $\pm$ 0.053 	& 0.331 $\pm$ 0.058 	& 26.5$^{+4.3}_{-2.2}$ \\
	     & 0.432 $\pm$ 0.066 	& 0.269 $\pm$ 0.078 	& 24.8$^{+3.2}_{-3.0}$ \\
    	     & 0.255 $\pm$ 0.044 	& 0.0912 $\pm$ 0.0567 	& 6.5$^{+5.7}_{-3.7}$ \\
	     & 0.257 $\pm$ 0.032 	& 0.0933 $\pm$ 0.0385 	& 14.3$^{+5.9}_{-7.3}$ \\

Mrk\,817     & 0.75 $\pm$ 0.10 		& 0.448 $\pm$ 0.114 	& 19.0$^{+3.9}_{-3.7}$ \\
	     & 0.61 $\pm$ 0.06 		& 0.308 $\pm$ 0.062 	& 15.3$^{+3.7}_{-3.5}$ \\
	     & 0.612 $\pm$ 0.033 	& 0.310 $\pm$ 0.038 	& 33.6$^{+6.5}_{-7.6}$ \\

3C\,390.3    & 0.87 $\pm$ 0.14 		& 0.275 $\pm$ 0.192 	& 23.6$^{+6.2}_{-6.7}$ \\

PG\,2130+099 & 2.85 $\pm$ 0.26 		& 2.24 $\pm$ 0.27 	& 158.1$^{+29.8}_{-18.7}$ \\
\enddata
\tablenotetext{a}{Continuum luminosity measurements are taken from 
                  Table~1 of \citet{kaspi05}.}
\tablenotetext{b}{Galaxy contributions have been subtracted from 
                  corrected luminosities, and Galactic extinction
                  corrections have been applied as described in the
                  text.}
\tablenotetext{c}{BLR radii measurements are calculated using only the 
                  H$\beta$ line in the rest frame of the AGN and are 
                  taken from Table~6 of \citet{peterson04}.}      
\tablenotetext{d} {This measurement was deemed unreliable by 
                   \citet{peterson04}.}
\end{deluxetable}

\clearpage

\begin{deluxetable}{lccccccccccc}
\tablecolumns{12}
\tablewidth{0.0pt}
\tabletypesize{\scriptsize}
\rotate
\tablecaption{BLR \rl\ Fits}
\tablehead{
\colhead{} &
\colhead{} &
\colhead{} &
\multicolumn{3}{c}{FITEXY} &
\colhead{} &
\multicolumn{2}{c}{BCES} &
\colhead{} &
\multicolumn{2}{c}{GaussFit}\\ \cline{4-6} \cline{8-9} \cline{11-12}
\colhead{Note} &
\colhead{N\tablenotemark{a}} &
\colhead{} &
\colhead{K} &
\colhead{$\alpha$} &
\colhead{Scatter\tablenotemark{b}} &
\colhead{} &
\colhead{K} &
\colhead{$\alpha$} &
\colhead{} &
\colhead{K} &
\colhead{$\alpha$}}

\startdata
\multicolumn{12}{c}{H$\beta$ time lag only} \\ \hline 
A & 55 && $-19.9 \pm 2.3$ & $0.486 \pm 0.052$ & 45 && $-22.5 \pm 2.6$ & $0.546 \pm 0.059$ && $-20.3 \pm 1.8$ & $0.496 \pm 0.041$ \\
  & 32 && $-22.8 \pm 3.2$ & $0.551 \pm 0.072$ & 54 && $-23.7 \pm 3.7$ & $0.573 \pm 0.082$ && $-21.1 \pm 2.4$ & $0.514 \pm 0.055$ \\
B & 52 && $-19.4 \pm 1.6$ & $0.476 \pm 0.037$ & 31 && $-20.0 \pm 1.9$ & $0.491 \pm 0.044$ && $-19.5 \pm 1.4$ & $0.480 \pm 0.032$ \\
  & 29 && $-20.4 \pm 2.2$ & $0.499 \pm 0.051$ & 38 && $-20.1 \pm 2.6$ & $0.493 \pm 0.059$ && $-19.6 \pm 1.8$ & $0.482 \pm 0.041$ \\
C & 49 && $-19.9 \pm 1.5$ & $0.487 \pm 0.035$ & 26 && $-21.3 \pm 1.5$ & $0.520 \pm 0.034$ && $-20.3 \pm 1.4$ & $0.496 \pm 0.031$ \\
  & 26 && $-21.9 \pm 1.9$ & $0.530 \pm 0.043$ & 30 && $-22.9 \pm 2.2$ & $0.555 \pm 0.050$ && \boldmath{$-21.2 \pm 1.7$} & \boldmath{$0.518 \pm 0.039$} \\

\hline 
\multicolumn{12}{c}{Mean Balmer-lines time lag} \\ 
\hline 

D & 58 && $-20.4 \pm 2.2$ & $0.496 \pm 0.049$ & 47 && $-23.1 \pm 2.7$ & $0.560 \pm 0.060$ && $-20.8 \pm 1.8$ & $0.509 \pm 0.041$ \\ 
  & 35 && $-23.0 \pm 2.9$ & $0.554 \pm 0.066$ & 55 && $-24.4 \pm 3.6$ & $0.588 \pm 0.082$ && $-21.6 \pm 2.4$ & $0.526 \pm 0.055$ \\ 
E & 54 && $-19.7 \pm 1.6$ & $0.483 \pm 0.036$ & 34 && $-20.1 \pm 1.9$ & $0.494 \pm 0.044$ && $-19.9 \pm 1.4$ & $0.489 \pm 0.033$ \\ 
  & 31 && $-20.6 \pm 2.2$ & $0.502 \pm 0.050$ & 41 && $-20.2 \pm 2.7$ & $0.494 \pm 0.060$ && $-20.0 \pm 1.9$ & $0.490 \pm 0.043$ \\
C & 51 && $-20.4 \pm 1.4$ & $0.498 \pm 0.033$ & 27 && $-21.6 \pm 1.5$ & $0.526 \pm 0.034$ && $-20.5 \pm 1.3$ & $0.502 \pm 0.030$ \\ 
  & 28 && $-22.5 \pm 1.9$ & $0.546 \pm 0.041$ & 31 && $-23.2 \pm 2.1$ & $0.561 \pm 0.047$ && $-21.5 \pm 1.7$ & $0.524 \pm 0.038$ \\ 
 
\enddata
\vspace{-10pt}
\tablenotetext{a}{N is the number of pairs of \rl\ measurements included in 
		  each fit.}
\tablenotetext{b}{Scatter is given as percent of the measurement value of $r$.}
\tablecomments{ Two rows are given for each set of data: the first row gives 
             	the fit results where multiple data sets for each object are 
		treated individually, and the second row gives the fit results 
		where multiple measurements are averaged together as described
		in the text. The fit values in bold face are the fit to the 
		\rl\ relationship that is quoted throughout this paper. 
		{\bf A.} All reliable measurements for the
		reverberation-mapped AGN are included in this fit.  Only
		the objects PG\,0844+349, PG\,1211+143, and NGC\,4593 do
		not have a reliable H$\beta$ time lag measurement.
		{\bf B.} In addition to any objects that were excluded
		above, the low luminosity objects NGC\,3516, IC\,4329A,
		and NGC\,7469 have been excluded from this fit, for
		reasons described in the text.  		
		{\bf C.} NGC\,3227, NGC\,4051, and PG\,2130+099, as well
		as objects mentioned above, have been excluded from this
		fit for reasons described in the text.
		{\bf D.} All reverberation-mapped AGN were included in
		this fit. 
		{\bf E.} NGC\,4593, NGC\,3516, IC\,4329A, and NGC\,7469
		are excluded from this fit for reasons described in the
		text.  }
\end{deluxetable}

\clearpage

\begin{figure}
\figurenum{1}
\epsscale{0.5}
\plotone{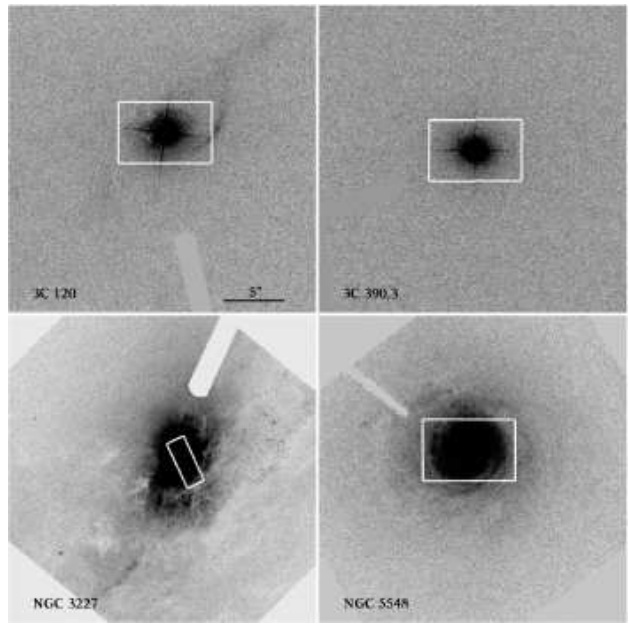}
\caption{ACS HRC images of four representative galaxies from our sample: 
	 3C\,120, 3C\,390.3, NGC\,3227, and NGC\,5548.  Each galaxy is
	 overlaid with the typical spectral aperture used in its
	 original monitoring campaign, centered on the position of the
	 AGN.  North is up and east is to the left.  The scale is the
	 same for each of the images.}
\end{figure}

\clearpage

\begin{figure}
\figurenum{2}
\epsscale{0.5}
\plotone{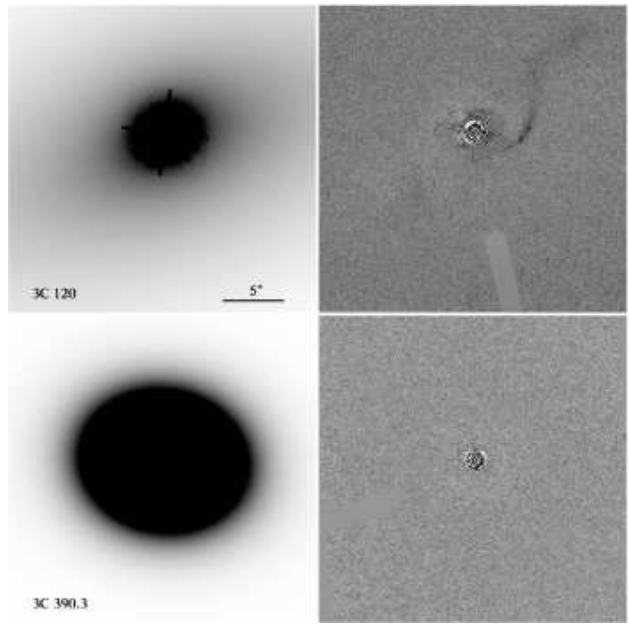}
\caption{Galaxy+PSF models and residuals for four representative galaxies 
	 from our sample: 3C\,120, 3C\,390.3, NGC\,3227, and NGC\,5548.
	 North is up and east is to the left.  The scale is the same for
	 each of the images.}
\end{figure}

\clearpage

\plotone{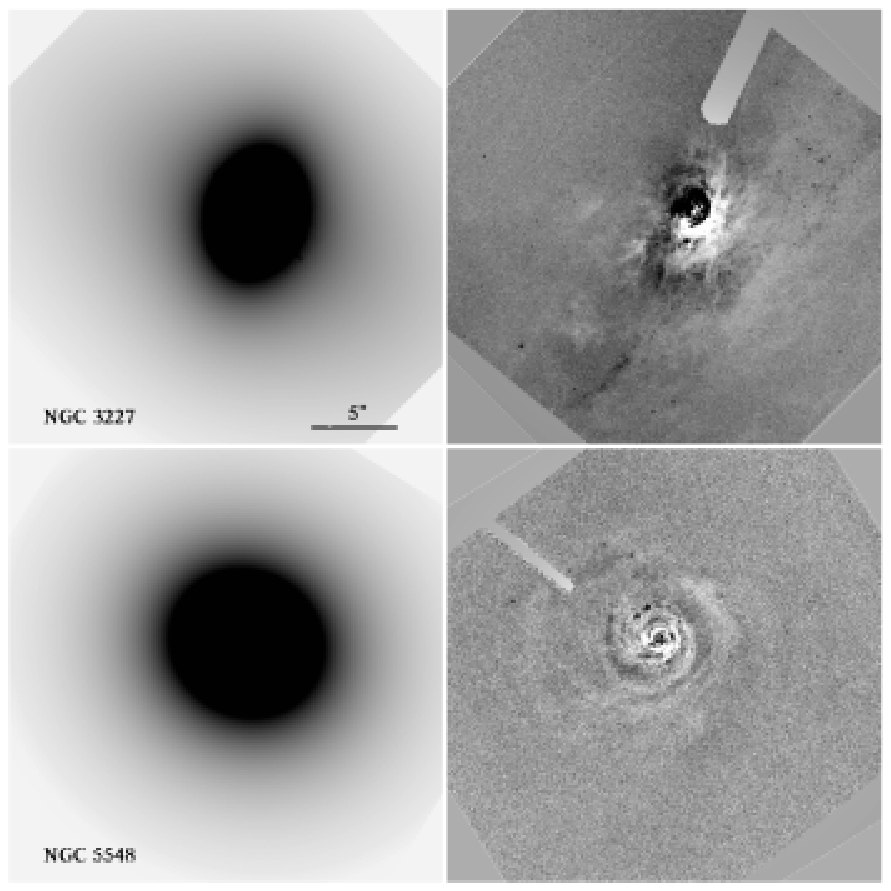}
\centerline{Fig. 2. --- continued.}

\clearpage

\begin{figure}
\figurenum{3}
\epsscale{0.5}
\plotone{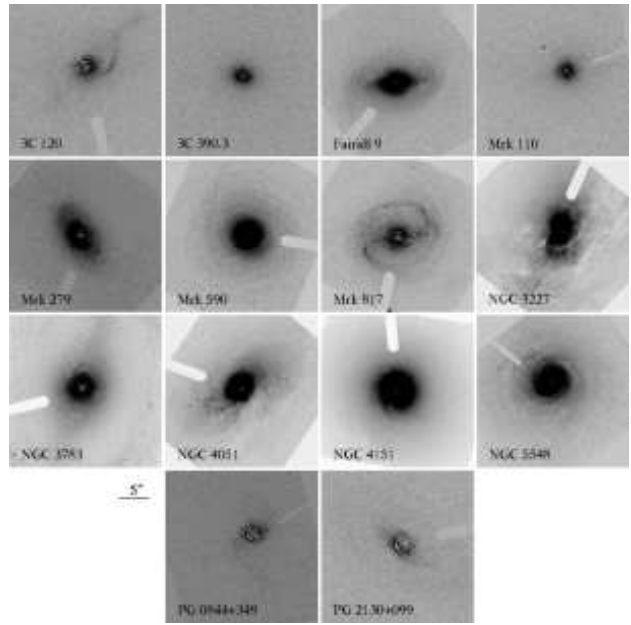}
\caption{PSF-subtracted images for all 14 galaxies in this study.  North 
	 is up and east is to the left.  The scale is the same for each of
	 the images.}
\end{figure}

\clearpage

\begin{figure}
\epsscale{1}
\figurenum{4}
\plotone{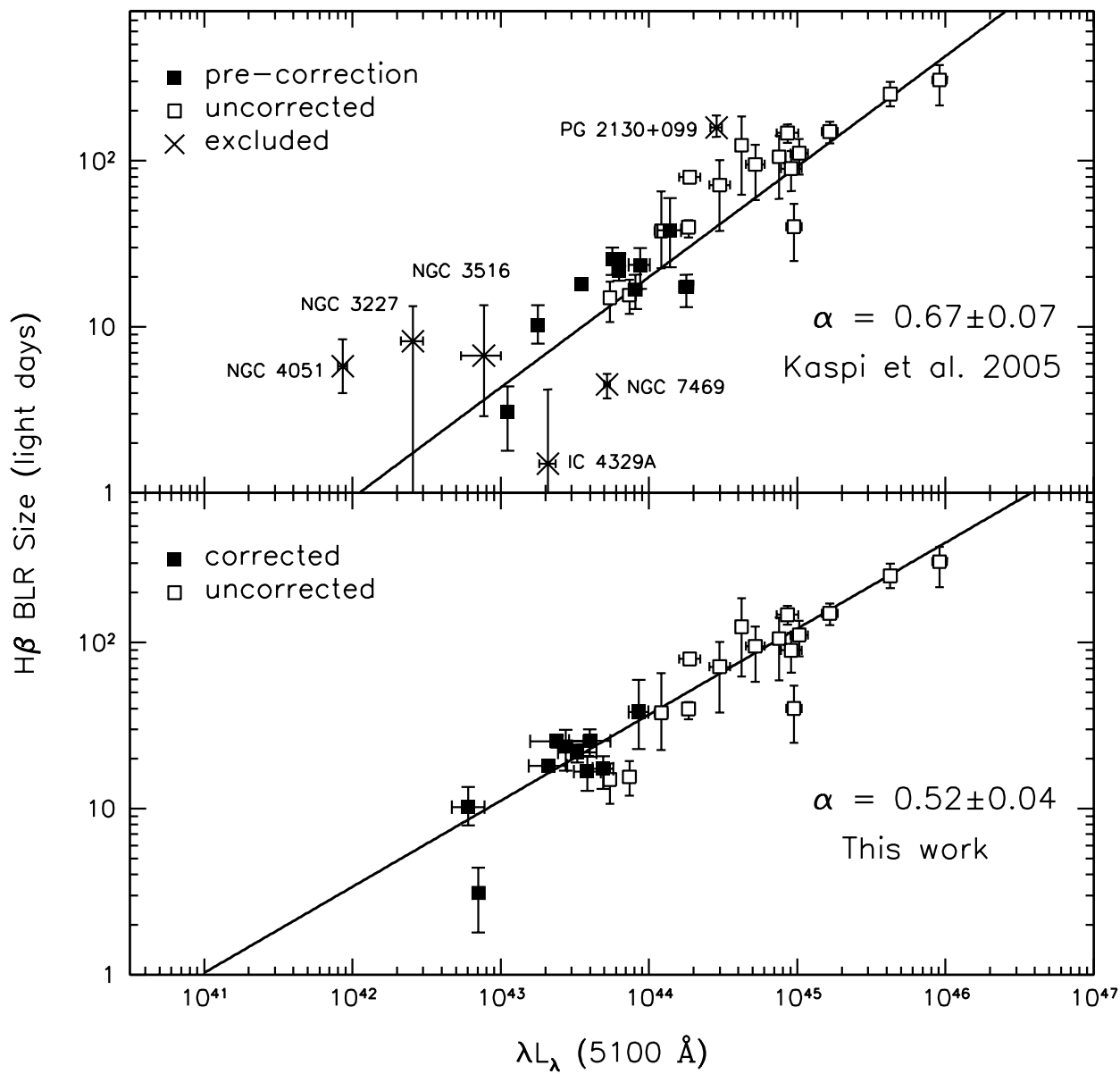}
\caption{H$\beta$ BLR size versus the luminosity at 5100\,\AA\ for the 
	reverberation-mapped AGN.  Open boxes are from
	\citet{kaspi05}. Filled boxes are also from \citet{kaspi05}, but
	are corrected for the host galaxy starlight contribution.  The
	top fit is the power-law fit determined by \citet{kaspi05} and
	the bottom fit is the power-law fit determined in this work.
	The crosses show the locations of objects that were excluded
	from various fits for reasons described in the text.}
\end{figure}

\end{document}